\documentstyle[preprint,aps,amstex,epsfig]{revtex}

\begin{document}
\draft
\title
{Efficient Diagonalization of Kicked Quantum Systems} 
\author{R. Ketzmerick,$^{1,2}$
K. Kruse,$^{2}$ and
T. Geisel$^{1,2}$}

\address{
$^1$Institute for Theoretical Physics, 
University of California, Santa Barbara, CA 93106, USA \\
$^2$Max-Planck-Institut f\"ur Str\"omungsforschung und Institut f\"ur
Nichtlineare Dynamik der Universit\"at G\"ottingen, 
Bunsenstra{\ss}e 10, D-37073 G\"ottingen, Germany$^*$\bigskip\bigskip\\  
\parbox{14cm}{\rm  
We show that the time evolution operator of kicked quantum systems,
although a full matrix of size $N\times N$, can be diagonalized with
the help of a new method based on a suitable combination of Fast
Fourier Transform and Lanczos algorithm in  
just $O\!\left(N^2\ln N\right)$ operations. It allows the diagonalization of
matrizes of sizes up to $N\approx 10^6$ going far beyond the
possibilities 
of standard diagonalization techniques which need $O\!\left(N^3\right)$
operations. We have applied this method to the 
kicked Harper model revealing its intricate spectral properties.
\smallskip\\
PACS numbers: 02.60.Dc, 05.45.+b
}}

\maketitle
\narrowtext

\section{Introduction}

Periodically kicked quantum systems have played a prominent role 
in studies of quantum chaos~\cite{ccif} for the analysis of
signatures of classical chaos in quantum
systems~\cite{haake}. Important examples are the 
kicked rotator, i.e., the quantum version of Chirikov's standard
map~\cite{ccif,izrailev}, the kicked top~\cite{haake}, and the 
kicked Harper model~\cite{artuso3}. Their quantum time evolution
operator allows a fast numerical iteration on wave packets
just like in their classical analogs, which are maps that can be iterated
very quickly. The study of 
kicked sytems has led to many discoveries, a very spectacular one being
the suppression of classical chaotic diffusion due to quantum
mechanical interference in the kicked rotator~\cite{ccif}, which later on
was understood by a mapping onto the Anderson model of disordered
systems~\cite{fgp} and onto the supersymmetric nonlinear
$\sigma$ model for quasi one-dimensional wires~\cite{az}. 
Another example supporting the importance of kicked
systems appeared in the study of quantum signatures of classical
chaos in quasiperiodic systems, in which the kicked Harper 
model~\cite{artuso3,leboeuf1,lima,geisel,ketzmerick,artuso1,artuso2,guarneri,roncaglia,leboeuf2,dana,borgonovi,ketzmerick1}
yielded surprising
phenomena~\cite{lima,artuso1,artuso2,borgonovi,ketzmerick1}, 
e.g., metal-insulator transitions tuned by classical chaos.
In the future, kicked quantum systems most probably
will be the first to be used for numerical studies of the quantum signatures
of mixed systems having a hierarchical phase space structure at
the boundary between regular and chaotic motion. For example, the
semiclassically  predicted~\cite{rk96} and experimentally
observed~\cite{hegger,sachrajda} fractal properties of conductance
fluctuations wait to be verified by numerical quantum calculations.

In this paper we will show how the Lanczos algorithm for diagonalizing
matrices \cite{cw}, which is
well suited for {\it sparse} matrices, can be used to exploit the
advantages of kicked quantum systems even though their time evolution
operator is a {\it full} matrix. 
The key point, which has been overlooked throughout twenty
years of studying quantum chaos with kicked systems, is the following
observation: The 
most time consuming step of the Lanczos algorithm, a matrix-vector
multiplication, is analogous to a step of the
time evolution for some wave packet. For kicked systems the time
evolution is known to be efficiently performed 
by using the Fast Fourier Transform (FFT)
algorithm~\cite{numrec}. Consequently, by 
combining the Lanczos algorithm and the FFT, diagonalization of kicked
systems 
only takes of the order of $N^2\ln N$ operations and for a matrix of size
$N=10^5$ needs about two days on a standard 
workstation. By using parallel
computers matrices of size $N=10^6$ should be diagonalizable. 

As a first application of this new method, we will study the quantum
signatures of classical chaos in quasiperiodic systems. For the kicked
Harper model (KHM) {\em off} the critical line, Borgonovi and
Shepelyansky~\cite{borgonovi} convincingly concluded from
diagonalizing matrices of sizes up to $N=1775$ that in the limit
$N\rightarrow\infty$ the spectrum, in addition to pure point and
absolutely continuous components, can also have a singular continuous
component. The Lanczos-FFT method allows us to easily diagonalize the
KHM for the much larger size of $N=51536$. We will show that the
indications for a singular continuous component found in Ref.~\cite{borgonovi}
do not persist for larger system sizes. We will thereby
reveal the KHM's intricate spectral properties and demonstrate the 
applicability of our method.

\section{The Lanczos-FFT method}

The numerical advantage of the time evolution of a periodically
kicked system, e.g., a one-dimensional Hamiltonian of the form
\begin{equation}\label{hamkick}
H\left( t\right) = T\left(p\right) +
V\left( x\right)\sum_n \delta\left( t-n\right) ,
\end{equation}
is due to the fact, that its
time-evolution operator for one period 
factorizes into
\begin{equation}\label{fakt}
U=\exp\left\{-\frac{i}{\hbar}T\left(p\right)\right\}
  \exp\left\{-\frac{i}{\hbar}V\left(x\right)\right\}.
\end{equation}
Here, $\hbar$ is typically an {\em effective} Planck's constant, $p$ is the
momentum and $x$ the position operator. 
The second factor describes the kick, whereas the first factor governs
the free evolution between kicks. 
If $U$ is periodic in $x$, then $U$ can be represented as an
infinite matrix depending on a Bloch phase $\theta_x$, the so-called
quasi-momentum. If, in addition, $U$ is periodic with period  
$N$, i.e., $U_{l+N,l^\prime+N}=U_{l,l^\prime}$, the time-evolution
operator can be reduced to an $N \times N$ matrix depending on a
second Bloch phase $\theta_p$. This is, e.g., the case for the kicked
rotator and the kicked Harper model, when $\hbar$ takes the value
$2\pi M/N$. In order to obtain the time evolution it will
be applied iteratively to a wave packet described by a vector of
length $N$. This application in general requires $N^2$ operations for
a single time step, whereas the factorization of $U$
[Eq.~(\ref{fakt})] allows a much faster 
implementation: In either momentum or position representation,
application of the first and second factor of $U$, resp., amounts to a
vector-vector multiplication using just $N$ operations. Going back and
forth between position and momentum representation one uses the
very efficient FFT algorithm, which needs
of the order of $N\ln N$ operations~\cite{numrec}. A detailed
description of this standard method is given in the appendix
including its extension to cases where $N$ is not a power of 2. Thus, for
propagating a wave packet one period in time just $O\left(N\ln
N\right)$ operations are needed, which allows the study of time
evolutions in systems with sizes up to $N \approx 10^6$.

On the other hand, this dramatic advantage of kicked quantum systems
when studying time 
evolutions has not yet been fully exploited for
determining the {\em eigenvalues} of $U$. So far this
advantage has been used in this context to perform the time evolution
for $zN$ steps 
taking just $zN^2\ln N$ operations and to Fourier transform the
resulting time series giving the unitary eigenvalues. Unfortunately, for large
matrices this method fails for two reasons:
(i) resolution: very often, the finite resolution $1/(zN)$ will not
resolve closeby eigenvalues. Even worse, for quasiperiodic quantum systems
with fractal spectra of dimension $D<1$, in which the typical level
spacing scales as $1/N^{1/D}$,
this requires at least $O\left(N^{1+1/D}\ln N\right)$ operations.
(ii) storage: if $U$ has localized eigenfunctions, one needs to store the
time evolution at $O\!\left(N\right)$ sites and needs storage of 
$O\!\left(z N^2\right)$ bytes, which 
for $N=10^6$ and a relatively small $z=10$ amounts to $10^5$ Gigabyte
by far exceeding conventional storage capacities. 
Standard diagonalization methods need $O\left(N^3\right)$
operations and have so far restricted $N$ to values of only a few
thousand.

We will now describe explicitly the algorithm which makes use of the
factorization of $U$ 
[Eq.~(\ref{fakt})] to obtain all eigenvalues in $O\!\left(N^2\ln
N\right)$ operations.
To this end we start by recalling the
standard Lanczos algorithm for a general  
$N\times N$ complex Hermitian matrix $H$~\cite{cw}. It generates a
sequence of {\em tridiagonal} real symmetric matrices 
\begin{equation}
L_n=\left(\begin{array}{cccc}
\alpha_1\,\, & \beta_2\,\, \, &         &          \\
\beta_2\,\,  & \alpha_2\,\,\, & \ddots\,\,\,  &          \\
         & \ddots\,\,\,   & \ddots\,\,\,  &\, \beta_n  \\
         &          & \beta_n\, \,\,& \,\alpha_n \\
\end{array}\right),
\end{equation}
\noindent whose eigenvalues can be calculated very efficiently by 
standard routines \cite{numrec} and converge towards the eigenvalues
of $H$. The calculation of $\alpha_n$ and $\beta_n$ is based on the following
recursion relations: Let $\psi_0, \psi_1 \in{\mathbb C}^N$ with
$\psi_0 \equiv 0$ and $\psi_1$ a normalized random vector.
Then for $n\ge 1$ we have 
$\alpha_n = \left(\psi_n,H\psi_n\right)$, $\beta_n =
\left(\psi_n,H\psi_{n-1}\right)$, and
$\psi_{n+1}=\bar{\psi}_{n+1}/\|\bar{\psi}_{n+1}\|$, where $\bar{\psi}_{n+1} =
H\psi_n-\alpha_n\psi_n-\beta_n\psi_{n-1}$. If the
calculation is carried out in exact arithmetic, then $L_N$ has the
same eigenvalues as $H$. In fact, in this case $L_N$ is related to $H$
via a similarity transformation $\Psi$, i.e., $\Psi^{-1} H\Psi =L_N$
with $\Psi =\left(\psi_1 \ldots\psi_N\right)$. Due to the instabilty
of the algorithm for finite precision arithmetic, $n$ has to
be larger than $N$ in order to 
approximate all eigenvalues of $H$ with sufficient accuracy. In this
case, $L_n$ will possess spurious eigenvalues not approximating any
eigenvalue of $H$. These so-called ghosts can be detected efficiently
using standard techniques \cite{cw}. Even though this algorithm
is designed for Hermitian matrices it can be used to obtain the
spectrum of a unitary matrix $U$ by applying it to
$H^+=\frac{1}{2}\left(U+U^\dagger\right)$ and 
$H^-=\frac{1}{2i}\left(U-U^\dagger\right)$. Then, for an eigenvalue
${\rm e}^{i\omega}$ of $U$ the eigenvalues of $H^+$ and $H^-$ are
$\cos\omega$ and $\sin\omega$, resp.

The most time consuming step in each iteration of the Lanczos
algorithm is the calculation of $H\psi_n$ which in general needs $N^2$
operations. If $H\psi_n$ can be computed in fewer operations as is the
case for sparse matrices, the algorithm will become faster than
standard diagonalizaton techniques. As described above, for kicked
systems in which $U$ is a 
{\it full} matrix, one can make use of its factorization [Eq.(\ref{fakt})]
and the FFT algorithm to
calculate $U\psi_n$ and thus $H^\pm\psi_n$ in just $O\left(N\ln
N\right)$ operations. Hence, by using the Lanczos-FFT method the
eigenvalues of $U$ can be calculated in $O\left(N^2\ln N\right)$
operations, such that 
systems of sizes up to $N\approx 10^6$ are treatable.

Let us finally mention that the numerical workload may be further
reduced if $U$ obeys certain symmetries. i) If the $N\times N$ matrix $U$
is symmetric, then $H^+=\rm{Re} U$ and
$H^-=\rm{Im} U$ such that $U^\dagger$ no longer occurs. This is, e.g.,
the case for the kicked Harper model if one uses the time evolution operator
$U$ for one period starting halfway between two kicks with
$\theta_p=0$. The spectrum, of course, is independent of the chosen time
when the one period time evolution operator starts. ii) If  the
eigenvalues $\rm{e}^{i\omega}$ of $U$ are symmetric with respect to
$\omega=0$, it suffices to diagonalize $H^+$. Each of these symmetries
reduces the number of operations by a factor of 2.
This completes the description of our method to
determine the spectrum of a kicked quantum system in $O\left(N^2\ln N\right)$
operations. 

\section{The spectrum of the kicked Harper model}

We will now apply the Lanczos-FFT method to the kicked Harper model
(KHM)~\cite{artuso3,leboeuf1,lima,geisel,ketzmerick,artuso1,artuso2,guarneri,roncaglia,leboeuf2,dana,borgonovi,ketzmerick1}.
By calculating the quasienergies of a system of size $N\approx 50000$
we will give strong evidence 
that a former result on the spectral properties of the
KHM obtained by treating matrices of sizes up to $N\approx 2000$
cannot be maintained. Since kicked systems with a larger size can up
to now only be treated by the Lanczos-FFT method, this demonstrates
the value of this algorithm.

As a generalization of the Harper model~\cite{harper,azbel,hofstadter}
the KHM is widely used to study the influence of a chaotic
classical limit on the spectral properties of a quasiperiodic quantum
system~\cite{lima,artuso1,artuso2,borgonovi,ketzmerick1}. The Harper
model is described by a tight-binding Hamiltonian with potential
$V_n=V\cos\left(2\pi\sigma n+\nu\right)$ at site $n$. The Hamiltonian
of the KHM is given by 
\begin{equation}\label{KHM}
H = L\cos p + K\cos x \sum_n\delta\left( t-n\right),
\end{equation}
\noindent i.e., it is of the form (\ref{hamkick}) with $T\left(p\right) =
L\cos p$ and $V\left(x\right) = K\cos x$. For small values of $K$ and
$L$ its spectrum is closely related to the spectrum of the Harper
model~\cite{geisel}. To be explicit, for small
$L/\hbar$ and $K/\hbar$ the quasienergies of the KHM are given up to a
factor
$L/2\hbar$ by the eigenenergies of the Harper model with $V=2K/L$,
$\sigma=\hbar/2\pi$ and $\nu=\theta_x$. Therefore, in the case of
small $K$ and $L$ the spectral 
properties of the KHM are the same as for the Harper model: For a typical
irrational $\hbar/2\pi$ and fixed quasimomentum $\theta_x$ the spectrum is
pure point if $K>L$, absolutely continuous if $L>K$, and singular continuous
for $K=L$ with a fractal dimension given by that of the Harper model.
Due to a duality property of the KHM~\cite{bb} on the critical line
$K=L$, the spectrum 
remains singular continuous for all values of $K=L$. Its fractal dimension,
though, changes with increasing $K$~\cite{artuso1}. Off the critical line, for
increasing values of $K$ and $L$ the
spectral properties of the KHM differ from the Harper model 
in that in some regions pure point and absolutely continuous
components coexist~\cite{lima,artuso2}. 
The changes in the spectral type for $K\neq L$ can
be understood with the help of avoided band crossings
(ABCs)~\cite{ketzmerick1}. These are 
present in the spectrum of the KHM because its classical version
generates chaotic dynamics (see Fig.~5 in Ref.~\cite{ketzmerick1}). 
As has been shown in
Ref.~\cite{ketzmerick1} they may tune metal-insulator
transitions, thus explaining the coexistence of pure point and absolutely
continuous spectral components. Since the fractal dimensions of a spectrum may
also change due to ABCs~\cite{ketzmerick1}, they provide a common 
explanation for all the
changes in the quasienergy spectrum of the KHM mentioned above.

Based on a numerical analysis of matrices of sizes up to
$N=1775$, Borgonovi and Shepelyansky found indications that even a
singular continuous component should exist off the critical
line~\cite{borgonovi}. In 
particular, they have studied the case of $K=4$, $L=7$,
$\hbar=2\pi/(6+\sigma_G)$, where $\sigma_G=(\sqrt{5}+1)/2$ is the
Golden Mean. For these parameters they have calculated the
inverse participation ratio $\xi=\sum_n|\psi_n|^4$ of the
eigenfunctions $\psi$ for the approximants $2\pi55/419$, $2\pi144/1097$
and $2\pi233/1775$ of $\hbar=2\pi/(6+\sigma_G)$ yielding matrices of
sizes $N=419,1097$ and $1775$, resp. In some regions of
the spectrum they have found the inverse participation ratio of the
corresponding 
eigenfunctions to decrease as $N^{-1}$, indicating that there
the spectrum is absolutely continuous, while in other regions the
inverse participation ratio remains constant, indicating the spectrum
to be pure point. Most interestingly, for quasienergies in the
interval $[2.4,2.8]$ the 
inverse participation ratio of the corresponding eigenfunctions scaled
with some exponent larger than -1 and smaller than 0, indicating a
singular continuous spectral component in the limit $N \rightarrow \infty$.

Using the Lanczos-FFT method we have calculated the quasienergy spectrum
for the same values of $K$ and $L$ but for
$\hbar=2\pi 6765/51536$ that is a much higher approximant of
$2\pi/(6+\sigma_G)$. This yields a matrix of size $N=51536$,
by far exceeding the size of a matrix treatable by
standard methods. In Fig.~1 we show the quasienergy spectrum versus
the Bloch-phase $\theta_x$ and several magnifications of the quasienergy
interval $[2.4,2.8]$. Note that the isolated dots present in the figure
are ghosts produced by the Lanczos algorithm. 
From a plot of the eigenvalues of a periodic approximant as given
here, one can infer the spectral type of the quasiperiodic system as
follows: The spectrum is pure point if the width of Bloch bands is
negligible as compared to their spacings. It is absolutely continuous
if the spacings are negligible compared to the band widths. The
self-similar structure of a singular 
continuous spectrum leads to band widths and spacings of comparable
size as can be found, e.g., for the Harper model at the critical
point. In Fig.~1 on intermediate scales
the spectrum displays a self-similar structure in the quasienergy interval
$[2.4,2.8]$, 
confirming the findings of Ref.~\cite{borgonovi}. As the
final magnifications show, however, on smaller scales the
self-similarity vanishes in many places and there the spectrum consists of 
either levels or bands corresponding to pure point and absolutely continuous
components, resp. 
Note, however, that we also find small regions, where the self-similarity
does not end even for our large matrix size (not shown in Fig.~1). This
indicates that the fraction of self-similar scaling regions is decreasing
with matrix size and questions the occurrence of a singular continuous
component in the limit $N \rightarrow \infty$. How can one understand 
this surprising property of the KHM?

This self-similar structure of the spectrum on intermediate
scales can be understood with the help of avoided band crossings.
As already mentioned, ABCs tune metal-insulator
transitions in the KHM leading to the coexistence of pure point and
absolutely continuous components of the spectrum for $K\neq L$.
Numerical investigations show that these components do not overlap. Hence,
there will be a transistion manifold in the (quasienergy,$K$,$L$)-space 
separating pure point from absolutely continuous regions.
If this transition manifold, which may have a very
complicated shape, would coincide with a line of fixed $K$ and $L$
over some quasienergy region, one would have a singular continuous component
in this spectral region.
Unless there is a special symmetry present (like for $K=L$), however, this
is very unlikely to happen. Typically, a line of fixed $K$ and $L$ will have
many {\em intersections} with this manifold, yielding quasienergy regions
with either pure point or absolutely continuous spectrum.
Thus there is no singular
continuous component but the absolutely continuous and pure point
components can be intermingled in a complicated manner depending on the
shape of the transition manifold.
The closer one is to an intersection with the transition manifold, 
the finer will be the scale where the
self-similar looking spectrum decides for being pure point or absolutely
continuous.
For this reason the spectrum looks
self-similar on intermediate scales, whereas on smaller scales pure point and
absolutely continuous components show up. This is just
what can be seen in the quasienergy spectrum shown in Fig.~1.
This also explains why for some regions of the spectrum that are even closer
to such an intersection, we find self-similar structure on all scales of our
finite matrix. In the limit of an infinite matrix size the fraction of
self-similar scaling regions should tend to zero and there should be no
singular continuous component.

In order to go beyond this qualitative understanding, we plan for the future
to study quantitatively how this fraction decays. Also, it will be
interesting to map out the transition manifold in the 
(quasienergy,$K$,$L$)-space by studying the spectrum for large matrices. One
may wonder if it has a finite or an infinite number of intersections with a
line of fixed $K$ and $L$.

\section{Concluding remarks}

In this article we have presented a very powerful method for
diagonalizing kicked quantum systems of sizes up to $N\approx 10^5$ on
standard present day workstations. The necessity of diagonalizing
matrices of this size was then demonstrated by the example of the
KHM. Its intricate spectral structure was  revealed and found to have its
reason in the occurence of avoided band crossings. As was alluded
to in the introduction, the possibility of obtaining the
eigenvalues of large matrices is of great 
interest also in other areas of quantum chaos. Currently, our method
is applied to the 
investigation of quantum fractal eigenstates~\cite{cms97}. In the
future, some possible applications include the study of level
statistics for very large spacings, the search for
quantum signatures of the hierarchical phase space of mixed systems,
and the study of predictions for disordered systems via the kicked
rotator~\cite{az}.
We express our hope that the Lanczos-FFT method will play a useful
role in the field of quantum chaos.

This research was carried out partially during two of the authors' stay at
the Institute for Theoretical Physics, Santa Barbara. T.~G. and
R.~K. gratefully acknowledge the hospitality of the ITP and its members.
It was supported by the NSF under Grant No. PHY94-07194 and in
part by the Deutsche Forschungsgemeinschaft.

\section{Appendix}

In this appendix we give the explicit expressions usually employed for
calculating 
$U\psi$, when $U$ may be factorized according to Eq.~(\ref{fakt}). In,
e.g., momentum representation we have 
\begin{equation}\label{U}
U =
\exp\left\{-\frac{i}{\hbar}T\left(p\right)\right\}F_{p\leftarrow x}
\exp\left\{-\frac{i}{\hbar}V\left(x\right)\right\}F_{x\leftarrow p},
\end{equation}
\noindent with the effective Planck's constant $\hbar$
and $F_{p\leftarrow x}$ and $F_{x\leftarrow p}$ denoting the
Fourier transform from position to momentum space and vice versa,
resp. If the first and the third factor on the r.h.s.\ are periodic
with period $N$, 
the operator reduces to a matrix of size $N \times N$ depending on two Bloch
phases. Then, these factors are diagonal matrices with elements 
$\exp\left\{-\frac{i}{\hbar}T\left(\hbar(l+\frac{\theta_x}{2\pi})\right)\right\}$
and 
$\exp\left\{-\frac{i}{\hbar}V\left(2\pi (k+\frac{\theta_p}{2\pi})/N\right)\right\}$,
resp., in which $k,l=0,\ldots ,N-1$, and
$\theta_x,\theta_p\in\left[0,2\pi\right]$ are the 
Bloch phases in position and momentum representation, resp. The
Fourier transform $F_{p\leftarrow x}$ is the matrix
\begin{eqnarray}
\left(F_{p\leftarrow x}\right)_{lk} & = &
  \frac{1}{\sqrt{N}}{\rm e}^{-2\pi i\left(l+\frac{\theta_x}{2\pi}\right)
  {\displaystyle (}k+\frac{\theta_p}{2\pi}{\displaystyle )}/N}  \\
 & = & {\rm e}^{-i\theta_x\theta_p/2\pi N}\;\; 
       {\rm e}^{-i k\theta_x/N}\;\;
        \frac{1}{\sqrt{N}}{\rm e}^{-2\pi i kl/N}\;\;
       {\rm e}^{-i l\theta_p/N}\label{FT}
\end{eqnarray}
\noindent which factorizes in a scalar, in two diagonal matrices, and a
standard Fourier transform, which can be implemented using the FFT
algorithm. $\left(F_{x\leftarrow p}\right)_{kl}$ is simply the
conjugate complex of 
$\left(F_{p\leftarrow x}\right)_{lk}$ such that in the final
expression for $U$ [Eq.(\ref{U})] 
the first two terms of Eq.(\ref{FT}) cancel each other. 

The advantage of using
the FFT algorithm exists only if $N$ is a power of 2; otherwise it is less
efficient. In situations where $N$ is not a power of $2$, as
is typically the case for the kicked Harper
model where an irrational value of $\hbar/2\pi$ is approximated by a
rational $M/N$, one may use the 
following strategy: One replaces the Fourier
transforms in ${\mathbb C}^N$ by Fourier transforms in ${\mathbb
C}^{\tilde{N}}$ where $\tilde{N}>2N$ is a power of 2. Before the first
FFT one extends the vector of size $N$ to a vector of
size $\tilde{N}$ by adding zeros and after the second FFT one retains
only the first $N$ elements of the resulting vector. In addition, one
has to replace the diagonal matrix of size $N\times N$ with elements 
$\exp\left\{-\frac{i}{\hbar}V\left(2\pi(k+\frac{\theta_p}{2\pi})/N\right)\right\}$
by a diagonal matrix of size $\tilde{N}\times\tilde{N}$ with elements
\begin{equation}
\frac{1}{N}\sum_{j=-N+1}^{N-1}\sum_{m=0}^{N-1}
\exp\left\{-\frac{i}{\hbar}V\left(2\pi\Big(m+\frac{\theta_p}{2\pi}\Big)/N\right)\right\}
{\rm e}^{-2\pi i\;jl/N}
{\rm e}^{2\pi i\;jk/\tilde{N}},
\end{equation}
\noindent in which $k=0,\ldots,\tilde{N}-1$.

\begin{figure}[h]
\epsfig{figure=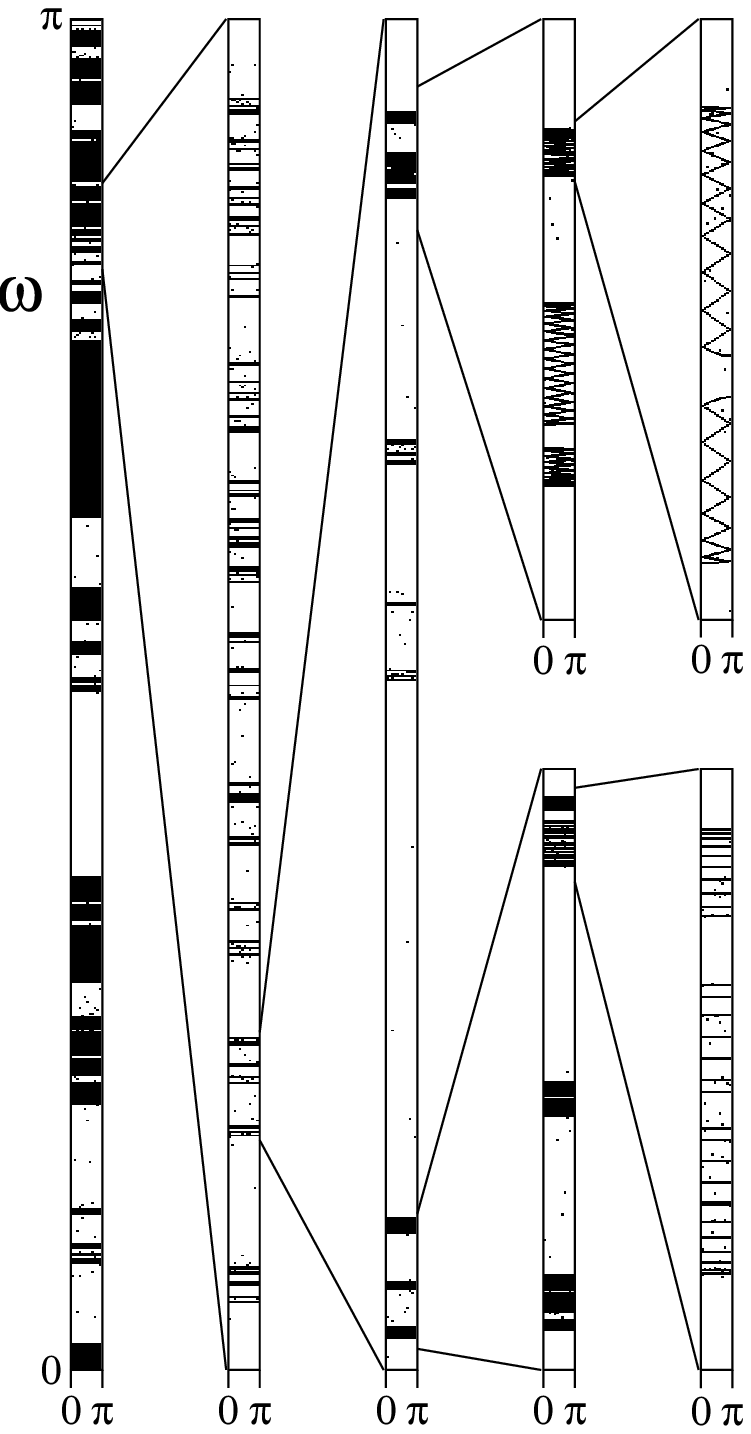,height=15.2cm,width=8.2cm}
\caption{Quasienergies $\omega$ vs.\ Bloch phase
$\theta_xN$ of the kicked Harper model [Eq.(\ref{KHM})] for $K\!=\!4$, 
$L\!=\!7$, $\hbar\!=\!2\pi\, 6765/51536$, and $\theta_p\!=\!0$.
Successive magnifications show fractal behaviour on
several scales, whereas the last magnification reveals either
localized eigenfunctions (no phase dependence of the quasienergies) or
extended eigenfunctions (strong phase dependence and no visible
gaps). The magnifications are for the intervals $[2.56,2.76]$,
$[2.594,2.61]$, $[2.6076,2.6092]$, $[2.60874,2.60891]$,
$[2.59425,2.59585]$ and $[2.59555,2.5958]$. The isolated points are 
ghosts due to the instability of the Lanczos algorithm.}
\end{figure}


\begin{references}
\bibitem[*]{address}
Present and permanent address.

\bibitem{ccif}
G.~Casati,\ B.~V.~Chirikov,\ F.~M.~Izraelev,\ and\ J.~Ford, in {\it
Stochastic Behavior in Classical and Quantum Hamiltonian Systems}
Vol.~93 of {\it Lecture Notes in Physics}, edited by C.~Casati and
J.~Ford (Springer, New York, 1979).

\bibitem{haake}
See e.g.,\ F.~Haake, {\it Quantum Signatures of Chaos}, Springer Series in
Synergetics, Vol. 54 (Springer-Verlag, Berlin Heidelberg, 1991);
G.~Casati and B.~Chirikov (eds.), {\it Quantum Chaos: Between order
and disorder} (Cambridge University Press, 1995).

\bibitem{izrailev}
F.~Izrailev, Phys.\ Rep. {\bf 196} (1990) 299.

\bibitem{artuso3}
R.~Artuso, G.~Casati, F.~Borgonovi, L.~Rebuzzini, and I.~Guarneri,
Int.\ J.\ Mod.\ Phys.\ B {\bf 8} (1994) 207. 

\bibitem{fgp}
S.~Fishman, D.~R.~Grempel, and R.~E.~Prange, Phys.\ Rev.\ Lett.\ {\bf
49} (1982) 509.

\bibitem{az}
A.~Altland and M.~R.~Zirnbauer, Phys.\ Rev.\ Lett.\ {\bf 77} (1996)
4536.

\bibitem{leboeuf1}
P.~Leb{\oe}uf, J.~Kurchan, M.~Feingold, and D.~P.~Arovas, Phys.\ Rev.\
Lett.\ {\bf 65} (1990) 3076. 

\bibitem{lima}
R.~Lima and D.~Shepelyansky, Phys.\ Rev.\ Lett.\ {\bf 67} (1991)
1377.

\bibitem{geisel}
T.~Geisel, R.~Ketzmerick, and G.~Petschel, Phys.\ Rev.\ Lett.\ {\bf
67} (1991) 3635. 

\bibitem{ketzmerick}
R.~Ketzmerick, G.~Petschel, and T.~Geisel, Phys.\ Rev.\ Lett.\ {\bf
69} (1992) 695. 

\bibitem{artuso1}
R.~Artuso, G.~Casati, and D.~Shepelyansky, Phys. Rev. Lett. {\bf 68}
(1992) 3826. 

\bibitem{artuso2}
R.\ Artuso, F.~Borgonovi, I.~Guarneri, L.~Rebuzzini, and G.~Casati,
Phys.\ Rev.\ Lett.\ {\bf 69} (1992) 3302. 

\bibitem{guarneri}
I.~Guarneri and F.~Borgonovi, J.\ Phys. {\bf A26} (1993) 119.

\bibitem{roncaglia}
R.~Roncaglia, L.~Bonci, F.~M.~Izrailev, B.~J.~West, and P.~Grigolini,
Phys.\ Rev.\ Lett.\ {\bf 73} (1994) 802.

\bibitem{leboeuf2} 
P.~Leb{\oe}uf and A.~Mouchet, Phys.\ Rev.\ Lett.\ {\bf 73} (1994)
1360.

\bibitem{dana}
I.~Dana, Phys.\ Rev.\ Lett.\ {\bf 73} (1994) 1609.

\bibitem{borgonovi}
F.~Borgonovi and D.~Shepelyansky, Europhys. Lett {\bf 29} (1995) 117.

\bibitem{ketzmerick1} R.~Ketzmerick, K.~Kruse, and T.~Geisel, Phys.\
Rev.\ Lett., {\bf 80} (1998) 137.

\bibitem{rk96} R. Ketzmerick, Phys. Rev. B {\bf 54} (1996) 10841.

\bibitem{hegger} H. Hegger et al., Phys. Rev. Lett. {\bf 77}, (1996) 3885.

\bibitem{sachrajda} A. S. Sachrajda et al., Phys.\ Rev.\ Lett.\ {\bf
80} (1998) 1948. 

\bibitem{cw}
J.~K.~Cullum\ and\ R.~A.~Willoughby, {\it Lanczos Algorithms
for Large Symmetric Eigenvalue Computations}, Progress in Scientific
Computing, Vol.~3 (Birkh\"auser, Boston, 1985).

\bibitem{numrec}
W.~H.~Press, S.~A.~Teukolsky, W.~T.~Vetterling, and B.~P.~Flannery,
{\it Numerical Recipes, 2nd ed.} (Cambridge University Press, 1992).

\bibitem{cms97}
G.~Casati, G.~Maspero, and D.~Shepelyansky, www site {\em
http://xxx.lanl.gov/abs/cond-mat/9710118}. 

\bibitem{harper}
P.~G.~Harper, Proc.\ Roy.\ Soc.\ London {\bf A68} (1955) 874.

\bibitem{azbel}
M.~Y.~Azbel, Sov.\ Phys.\ JETP {\bf 19} (1964) 634.

\bibitem{hofstadter}
D.~R.~Hofstadter, Phys.\ Rev. B {\bf 14} (1976) 2239.

\bibitem{bb}
J.~Bellissard and A.~Barelli, in {\it Quantum Chaos - Quantum
Measurement}, edited by P.~Cvitanovi\'c, I.~C.~Percival, and A.~Wirzba
(Kluwer Academic Publishers, Dordrecht 1992).

\end{references}
\end{document}